\documentclass[]{elsart}
\usepackage[]{epsfig}
\begin{document}

\begin{frontmatter}

\title{The Cornell potential in Lee-Wick inspired electrodynamics}

\author{Anais Smailagic 
}
\address{ INFN, Sezione di Trieste,\\
          Trieste, Italy}
\author{Euro Spallucci  \thanksref{s}}
\thanks[s]{e-mail : euro@ts.infn.it}
\address{Dipartimento di Fisica, Gruppo Teorico, Universit\`a di Trieste,
and INFN, Sezione di Trieste,\\
Trieste, Italy}

\begin{abstract}
In the seventies, Lee and Wick proposed an interesting modification of classical electrodynamics
 that renders it finite at the quantum level. At the classical level, this
modified theory leads to a regular linear potential at short distances while also reproducing the Coulomb potential 
at large distances. It is shown that a suitable modification
of the Lee-Wick idea can also lead to a linear  potential at large distances.
For this purpose, we study an Abelian
model that ``~simulates~'' the $QCD$ confining phase while maintaining the Coulomb behavior at short distances.
\\
This paper is organized in three parts. In the first part, we present a pedagogical
derivation of the static potential in the Lee-Wick model between two heavy test charges
using the Hamiltonian formulation. In the second part, we describe 
a  modification of the  Lee-Wick idea leading to the standard Cornell potential. In the third part,
we consider the effect of replacing a point-like charge with a smeared Gaussian-type source, 
that renders the electrostatic potential finite
as $r\to 0$. 

\end{abstract}

\end{frontmatter}

\section{Introduction}
 Classical Electrodynamics is an example of a very satisfactory theoretical description of electromagnetic phenomena
 which agrees with experimental data. Nevertheless, there are still some
underlying difficulties related to the point-like description of elementary charges. Even at the quantum level, 
this classical problem
remains in the form of divergent Feynman integrals. This requires an ``~\emph{ad hoc}~''  
procedure of ``~sweeping under the rug~'' the divergent quantities. It is  fair to notice that  the predictions
of Quantum Electrodynamics
are still in excellent agreement with the experimental results, as long as, a perturbative approach remains
valid. \\
 To overcome the problem of divergences, the Maxwell Lagrangian
needs suitable modifications in the strong field regime, while preserving the standard form in the weak-field
regime. One possible modification was proposed by Lee and Wick  \cite{Lee:1969fy,Lee:1970iw}. 
However, this modification has introduced  new problems into the theory. In particular, it led to a Yukawa-like
correction to the potential between charges. In detail, the combination of Coulomb and Yukawa terms
introduces a linearly rising potential at short distances while leaving the $1/r$ behavior at large
distances.\\
This is an interesting result because it can give hints to the different, still unsolved, problem of searching for
a linearly confining potential in a theory of strong interactions.\\
 Although the theory of strong interactions  has its widely accepted description 
in terms of non-Abelian gauge fields, so far, no one has been able to extract a confining potential
in this formulation.\\
In a phenomenological framework, it is customary to use the sum of an attractive Coulomb part and a linearly rising 
long-distance part to describe the
mass spectrum of heavy quark-anti quark bound states. This is known as the Cornell potential \cite{Eichten:1978tg}. \\
It is interesting to note  that the Lee-Wick model  has the same sum of the two potentials, but with short/large
distances regimes exchanged.\\
On the basis of this observation, one could consider a modification of the Lee-Wick model leading to
the Cornell potential between charges.\\
A first, modest,  attempt  to implement this idea will be limited to classical electrodynamics, where 
the technical intricacies  of non-Abelian gauge theories are absent. Any positive outcome in this direction will
be helpful  eventually to tackle the more complicated non-Abelian situation.\\
However, exchanging the two regimes, will reintroduce the problem of the short distances behavior of the Coulomb
potential. In order to finally solve this problem, we propose a smearing of the source compatible
with the finite size of any physical charge particle. This will remove the singularity at the origin, while not effecting
the linear long-distances behavior.\\
The paper is organized as follows: in Sect.(\ref{lwed}) we review the Hamiltonian formulation of the Lee-Wick
model. That formulation is a particularly useful way to factor out the static potential from the dynamical degrees of freedom. 
We than recover the Yukawa correction to the Coulomb potential and obtain the linear behavior at short distances.
We also propose an alternative  interpretation of the Lee-Wick potential as due to the polarization of the vacuum through
the distances dependent ( or ``~\emph{running}~'') dielectric constant.\\ 
In Sect.(\ref{corn}) we modify the Lee-Wick model in a way to exchange short/long distances regimes. This  leads
to the desired long distances ``~confinement~'' of charges. The price to pay, is the reintroduction
of a divergent  behavior at short distances. In the final part of this Section we cure the aforementioned singular
behavior by 
attributing a finite size to the charge source.\\
Finally, in the Appendix, we give a list of formulae used to obtain the results in this paper.

\section{Lee-Wick model}
\label{lwed}
Higher order derivative theories have been often proposed as a cure for the ultraviolet divergences in quantum field theories. 
A remarkable example is given by higher
order derivative gravity where quadratic curvature terms are instrumental in getting a renormalizable quantum theory. 
However, the presence of higher order derivatives usually violates the unitarity of the theory. In
the seventies, Lee and Wick \cite{Lee:1969fy,Lee:1970iw} proposed a way to reconcile UV finiteness and 
unitarity by claiming that the extra poles in the propagator describe unstable ( heavy )
particles. These particles eventually decay restoring the unitarity of the theory.\\
Lee-Wick electrodynamics is described by the higher derivative Lagrangian
density \footnote{ We use the metric signature convention $-+++$. }

\begin{equation}
L_{LW}= -\frac{1}{4}F_{\mu\nu}\left(\, 1 -\frac{\partial^2}{m^2}\,\right) F_{\mu\nu}  - eJ^{\mu} A_\mu
\label{lw}
\end{equation}

where $m$ is a new constant with a dimension of mass in natural units. In the literature, it is often referred to
 as the “ heavy photon mass ”. We consider
this terminology misleading since the theory is explicitly gauge invariant. A true photon mass stands
in front a quadratic term in $A_\mu$ which
is absent in the Lagrangian (\ref{lw}). A proper interpretation of $m$ is through the
introduction of a characteristic length $l_0 \equiv 1/m$ that indicates the distances
range $r< 1/m $, where the Lee-Wick correction dominates over the Maxwell term.\\
A well known problem in the covariant formulation of gauge theories is the
presence of propagating non-physical degrees of freedom. The usual way to
deal with these states, while maintaining covariance, is to introduce Fadeev-Popov ghosts.
Alternatively, one may give up explicit covariance in favor of exposing only the
physical degrees of freedom of the gauge field using the Hamiltonian formulation of
the theory. This is the approach we shall follow in this paper in order to extract the static, classical 
interaction energy between two test charges.\\
There are different ways to achieve this goal in the quantum
theory \cite{Accioly:2010js,Accioly:2011zz,Gaete:1998vr,Gaete:1999iy,Gaete:2007zn,Gaete:2007sj,Gaete:2009xf}. 
The Hamiltonian formalism will explicitly
factorize the static interaction from the gauge field's dynamical
degrees of freedom.\\
We start by writing down the Lagrangian for the Lee-Wick model (\ref{lw}) as:
{\small
\begin{eqnarray}
 L_{LW}&=&-\frac{1}{2}\left(\,\partial^0 A^i-\partial^i A^0\,\right) \left(\, 1 -\frac{\partial^2}{m^2}\,\right)
 \left(\,\partial_0 A_i-\partial_i A_0\,\right)+\nonumber\\
 && -\frac{1}{2}\left(\,\partial^k A^i-\partial^i A^k\,\right) \left(\, 1 -\frac{\partial^2}{m^2}\,\right)
 \left(\,\partial_k A_i-\partial_i A_k\,\right) - e \rho A_0 -e j^k A_k\nonumber\\
&&
\end{eqnarray}
}
Introducing the electric and magnetic fields  

\begin{eqnarray}
 && E^i\equiv \partial^0 A^i-\partial^i A^0= -\dot{A}^i + \partial^i A_0 \ ,\\
 && B_i \equiv \frac{1}{2} \epsilon_{ijk}\left(\,\partial^j A^k-\partial^k A^j\,\right)
\end{eqnarray}

the Lee-Wick Lagrangian (\ref{lw}) reads

\begin{equation}
  L_{LW}=-\frac{1}{2} E^i \left(\, 1 -\frac{\partial^2}{m^2}\,\right) E_i 
-\frac{1}{2} B^i \left(\, 1 -\frac{\partial^2}{m^2}\,\right) B_i - e \rho A_0 -e j^k A_k
\end{equation}

To write the corresponding Hamiltonian we define the canonically conjugate
momenta. One can see immediately that there is no kinetic term for $A_0$. Thus, only $A_i$ has
a conjugate momentum

\begin{equation}
 \Pi^i\equiv \frac{\delta L_{LW}}{\delta \partial_0 A_i}=-\left(\, 1 -\frac{\partial^2}{m^2}\,\right) E^i 
\end{equation}

By Legendre transforming $L_{LW}$ one finds the Hamiltonian
{\scriptsize
\begin{eqnarray}
&& H_{LW}=\left(\, E^i +\partial^i A_0\,\right)\left(\, 1 -\frac{\partial^2}{m^2}\,\right) E_i -L_{LW}\nonumber\\
       &&=\frac{1}{2} E^i \left(\, 1 -\frac{\partial^2}{m^2}\,\right) E_i 
-\frac{1}{2} B^i \left(\, 1 -\frac{\partial^2}{m^2}\,\right) B_i - A_0\left[\, e \rho -
 \left(\, 1 -\frac{\partial^2}{m^2}\,\right)\partial_i E^i   \,\right] -e j^k A_k\nonumber\\
&& \label{wlw}
\end{eqnarray}
}
In (\ref{wlw}) $A_0$ is simply a Lagrange multiplier enforcing the modified Gauss law:

\begin{equation}
 \frac{\delta H_{LW}}{\delta A_0}=0 \longrightarrow \left(\, 1 -\frac{\partial^2}{m^2}\,\right)\partial_i E^i=-e \rho
\label{glw}
\end{equation}

Equation (\ref{glw}) can be rewritten as  Gauss law with a “~running~” charge

\begin{equation}
 \partial_i E^i=-e\left(\,\frac{m^2}{-\partial^2 + m^2} \,\right)\, \rho\label{effe}
\end{equation}

This is the customary interpretation given in quantum field theory where the
interaction strength becomes scale dependent due to the vacuum polarization
effects. Taking this for granted, it seems more consistent to attribute the
strength variation to the dielectric properties of the medium, i.e. the vacuum,
rather than to the charge itself. For this purpose, it is worth recalling that one usually works
in a system of units where $1/4\pi \varepsilon_0= 1$. Restoring the dielectric constant, equation (\ref{effe})
is written as
 
\begin{equation}
 \partial_i E^i=-\frac{e}{ \varepsilon_0}\left(\,\frac{m^2}{-\partial^2 + m^2} \,\right)\, \rho
\equiv \frac{e}{ \varepsilon \left[\,\partial^2/m^2\,\right]} \label{gauss}
\end{equation}

In equation (\ref{gauss}) $e$ is a point-like charge, commonly described by the distributional density 
$\rho\left(\, \vec{r}\,\right) = e\,\delta\left(\,\vec{r}\,\right)$, while $\varepsilon$ is a modified “~effective
 dielectric constant~” accounting for the polarization effects through the substitution

\begin{equation}
 \frac{1}{\varepsilon_0}\longrightarrow \frac{1}{\varepsilon\left[\,\partial^2/m^2\,\right] }
\end{equation}

We now proceed to calculate the Coulomb-like potential $\phi\left(\,r\,\right)$ from (\ref{gauss})

\begin{eqnarray}
&& \vec{E}= -\nabla\phi\ ,\\
&& \phi\left(\,r\,\right)= -\frac{e}{\varepsilon_0}\int \frac{d^3k}{\left(\, 2\pi\,\right)^3}\frac{1}{\vec{k}^{\, 2}}
\frac{m^2}{\vec{k}^{\, 2}+m^2} e^{i\vec{k\cdot \vec{r}}}\ ,\nonumber\\
&&= -\frac{e}{\varepsilon_0}\int \frac{d^3k}{\left(\, 2\pi\,\right)^3}\left[\, \frac{1}{\vec{k}^{\, 2}}-
\frac{1}{\vec{k}^{\, 2}+m^2}\,\right]\, e^{i\vec{k\cdot \vec{r}}}\label{fourier}
\end{eqnarray}

To evaluate the second term  one can use the Schwinger parametrization:

\begin{equation}
 \frac{1}{\vec{k}^{\,2} + m^2}=\int_0^\infty ds e^{-s\left(\,\vec{k}^{\,2} + m^2\,\right)}
\end{equation}

and then  perform a Gaussian integration over $\vec{k}$. An additional rescaling 
$m^2 s\equiv \tau$ gives the integral

\begin{equation}
\longrightarrow \frac{1}{r} \int_0^\infty\frac{d\tau}{\tau^{1/2}} e^{-\tau}
e^{-m^2r^2/4\tau}\equiv  \frac{1}{r} I\left(\, m^2\,\right)
\end{equation}

The first integral in  (\ref{fourier}) is simply given by $I\left(\, m=0\,\right)$. The complete
potential is

\begin{eqnarray}
 \phi\left(\,r\,\right) &&=-\frac{e}{4\pi^{3/2}\varepsilon_0} \frac{1 }{r}\int_0^\infty\frac{d\tau}{\tau^{1/2}} e^{-\tau}
\left(\, 1 -e^{-m^2r^2/4\tau}\,\right)
\label{cpot}
\end{eqnarray}

Let us look at the asymptotic behavior of (\ref{cpot}).
In the small distances limit $mr<< 1$ one finds a linear term:

\begin{eqnarray}
 \phi\left(\,r\,\right) &&=-\frac{e}{4\pi^{3/2}\varepsilon_0} \frac{1}{r}\int_0^\infty\frac{d\tau}{\tau^{1/2}} e^{-\tau}
\left(\, 1 - 1 + \frac{m^2r^2}{4\tau}+ \dots\,\right) \nonumber\\
&& = \frac{em^2}{8\pi\varepsilon_0 } r +O\left(\, r^3\,\right)
\label{cy}
\end{eqnarray}

This result shows that the Lee-Wick modification regularizes the short-distances behavior of the, otherwise divergent,
electrostatic potential.\\
A relatively simple way to evaluate (\ref{cpot})  is to notice that the exponential function is quickly vanishing 
both at the lower and upper integration limits. 
This leaves only  a non-zero contribution in a narrow strip around its maximal value $\tau_0$. 
Therefore, the integral can be calculated by expanding the exponent around $\tau_0=mr/2$  as

\begin{equation}
 \int_0^\infty \frac{d\tau}{\tau^{1/2}} e^{-\tau-r^2m^2/4\tau}\equiv \int_0^\infty \frac{d\tau}{\tau^{1/2}} 
e^{-\omega\left(\,\tau\,\right)}\ ,\qquad \omega\left(\,\tau\,\right)\equiv \tau + \frac{m^2 r^2}{4\tau}
\end{equation}

\begin{equation}
 \omega^\prime\left(\,\tau\,\right)=0 \longrightarrow \tau_0= \frac{mr}{2}
\end{equation}

With the above definitions, one finds

\begin{eqnarray}
 \int_0^\infty \frac{d\tau}{\tau^{1/2}} e^{-\tau-r^2m^2/4\tau}
 &&= \frac{1}{\tau^{1/2}_0} e^{-\omega\left(\,\tau_0\,\right)}
\int_{-\infty}^\infty d\tau e^{-\omega^{\prime\prime}\left(\,\tau_0\,\right)\left(\, \tau -\tau_0\,\right)^2/2}
\nonumber\\
&&=\frac{1}{\tau^{1/2}_0} e^{-\omega\left(\,\tau_0\,\right)}
\left(\, \frac{2\pi}{\omega^{\prime\prime}\left(\,\tau_0\,\right)}\,\right)^{1/2}
\end{eqnarray}

The  potential (\ref{cy}) is found  to be

\begin{equation}
 \phi\left(\,r\,\right)= -\frac{e}{4\pi\varepsilon_0}\frac{1}{r}\left(\, 1 - e^{-mr}\,\right)
\label{cyend}
\end{equation}

The result (\ref{cyend}) shows that the  emerging potential is composed of  a Coulomb part and a Yukawa-like correction.

\begin{figure}
\begin{center}
\includegraphics[height=6cm]{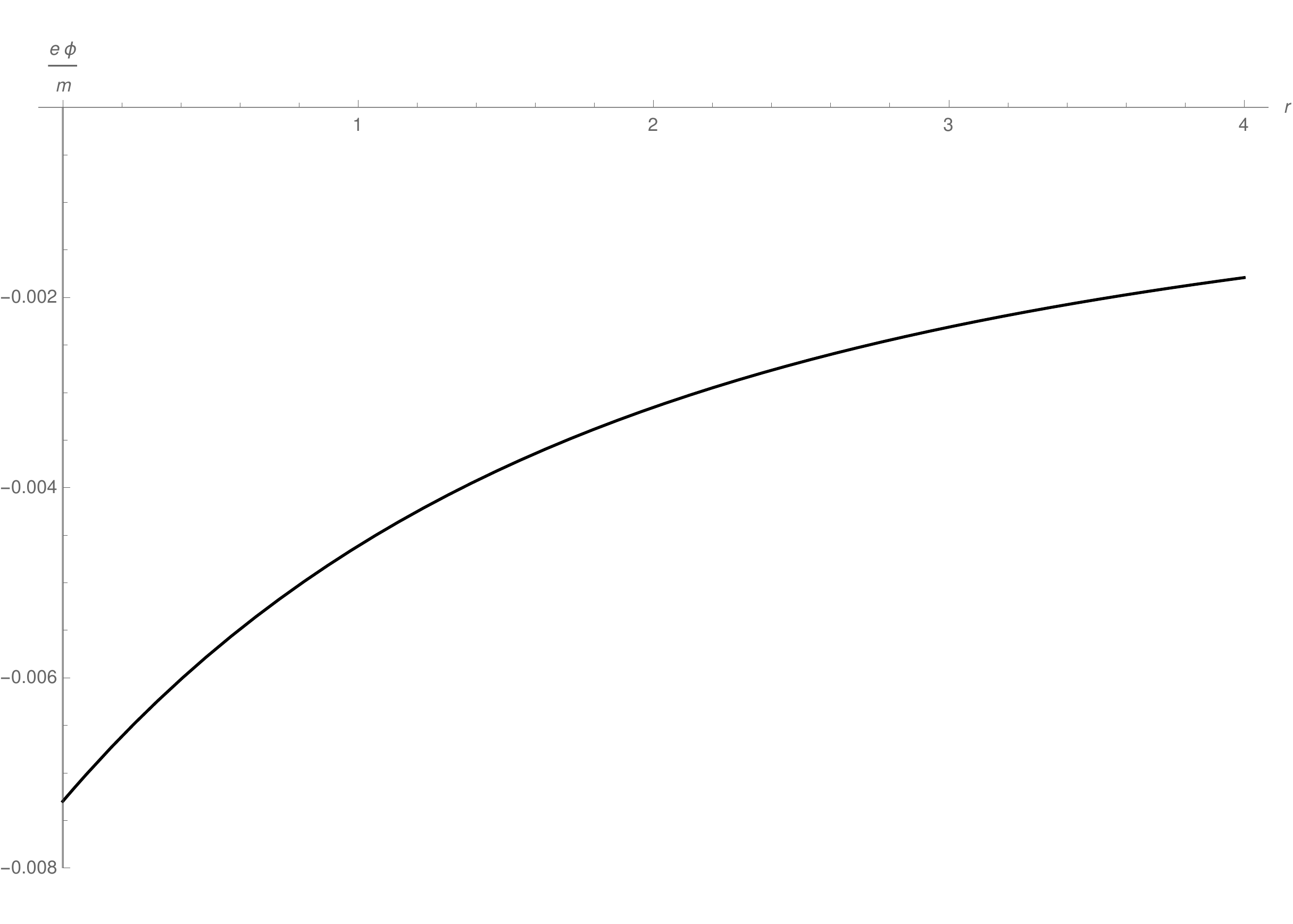}
\caption{Plot of the Coulomb-like potential (\ref{cpot}).}
\label{uno}
\end{center}
\end{figure}

Figure (\ref{uno})  shows that the potential is linear and finite near $r = 0$, while
approaches the Coulomb $1/r$ behavior at large distances. Accordingly, we see
that $\varepsilon\left(\,r\,\right)$ diverges as $r \to  0$ and tends to the usual value of the vacuum
dielectric constant $\varepsilon_0$ for $r >> 1/m$.

\begin{figure}[h!]
\begin{center}
\includegraphics[height=6cm]{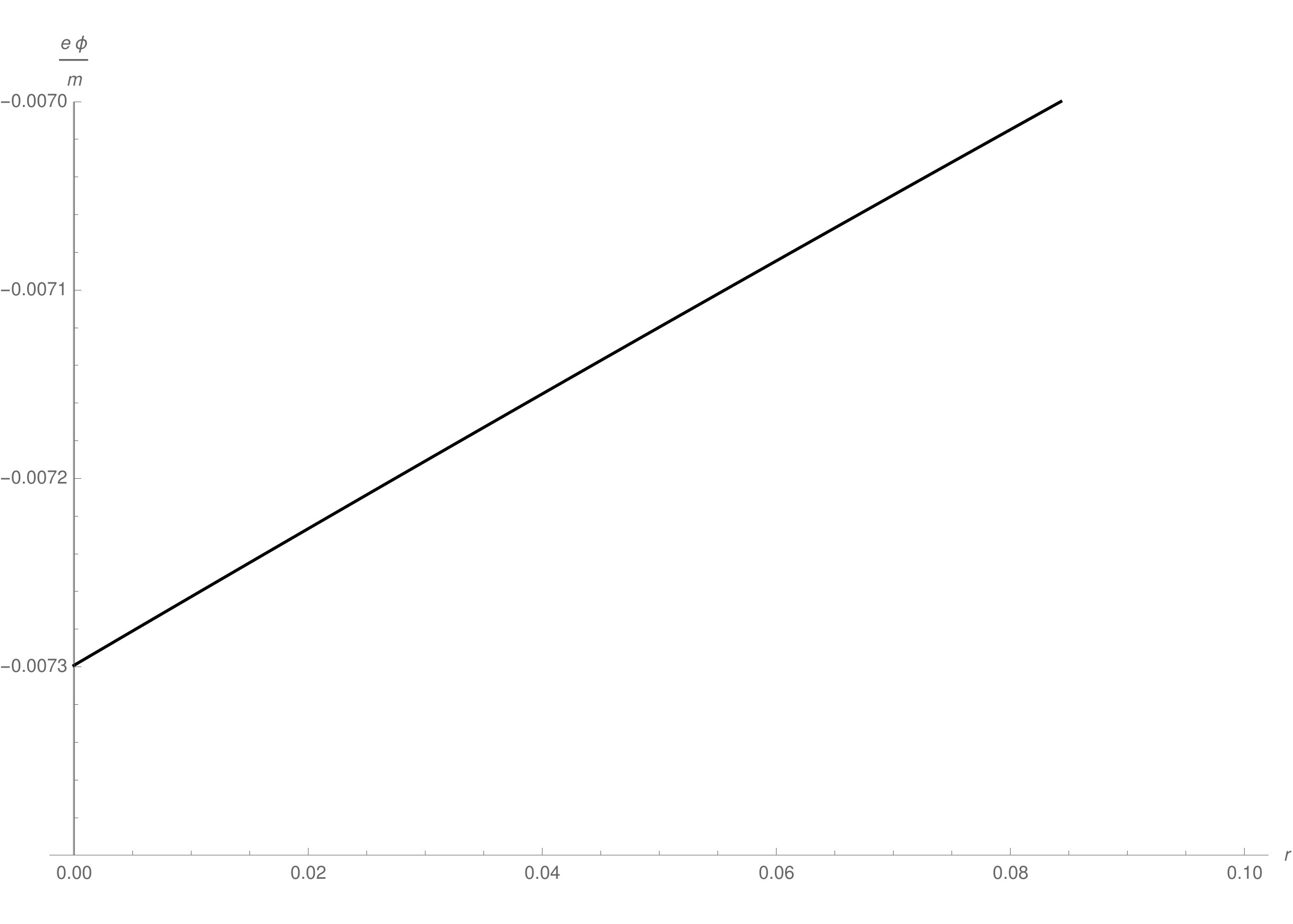}
\caption{Zoom of the Figure(\ref{uno}) near $r=0$, showing the linear behavior of the potential.}
\label{tre}
\end{center}
\end{figure}

Equation (\ref{cyend}) can  be rewritten in a Coulomb-like form with a position-dependent dielectric 
constant

\begin{equation}
 \phi\left(\,r\,\right) =\frac{e}{4\pi\varepsilon\left(\,r\,\right) \, r}\label{clike}
\ ,\qquad \varepsilon\left(\,r\,\right)=\frac{\varepsilon_0}{\left(\, 1 - e^{-m r}\,\right)}
\label{dielle}
\end{equation}

\begin{figure}[h!]
\begin{center}
\includegraphics[height=6cm]{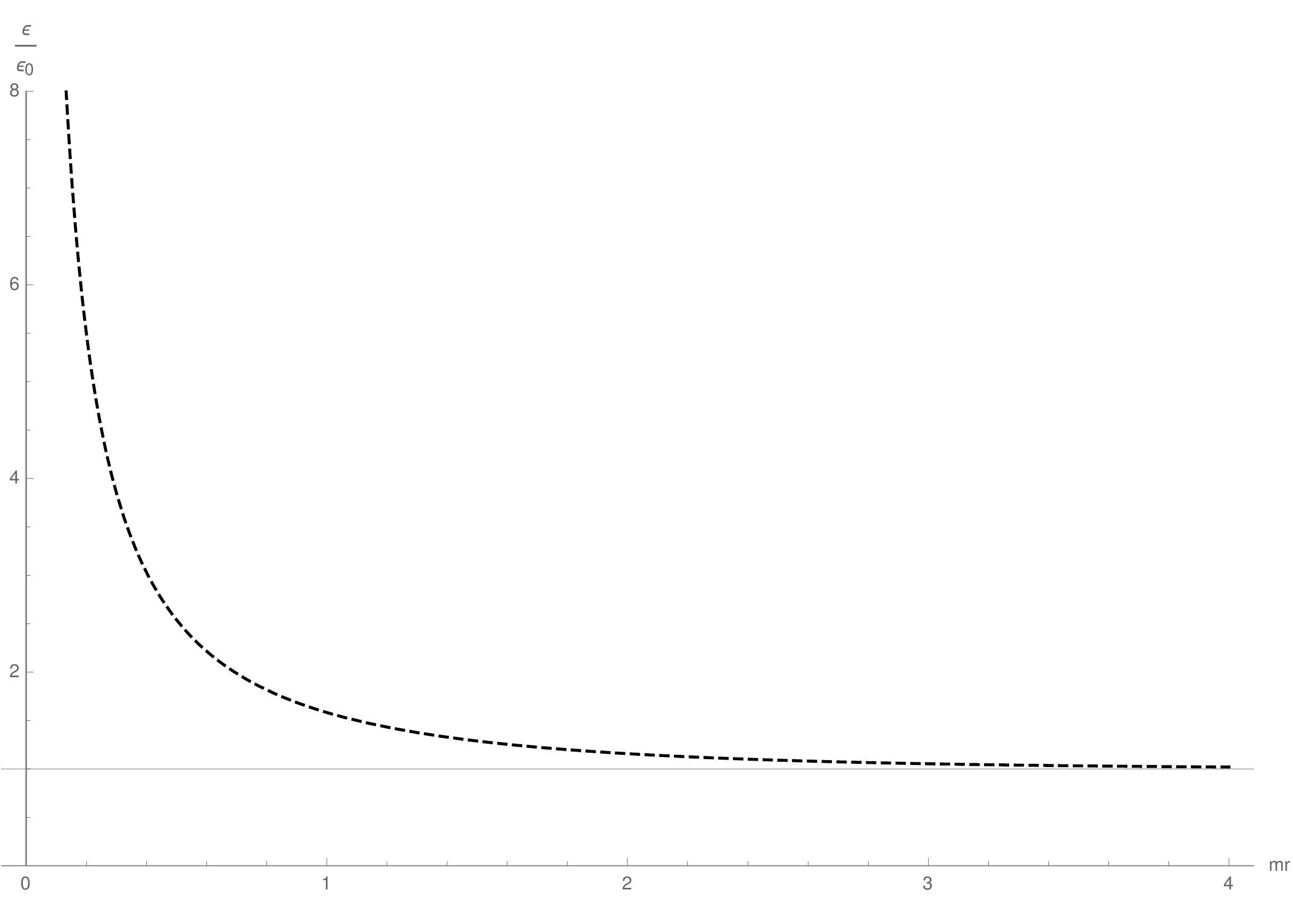}
\caption{Plot of the position-dependent relative dielectric constant (\ref{dielle}) .}
\label{due}
\end{center}
\end{figure}

This modified dielectric constant describes the ``~vacuum~'' polarization  induced by the test charge due to
virtual pair creation in its proximity. Figure (\ref{due}) below indicates an unbounded increase in 
$\varepsilon\left(\,r\,\right)$
as the distance decreases. ( This unphysical behavior of the dielectric constant is due to the choice of an idealized 
point-like charges. 
In the next subsection we shall improve this behavior by assuming a smeared ( non-point like ) source. )

\section{Cornell potential}
\label{corn}
In the previous Section we showed that it is possible to obtain a linear potential via a suitable modification of the 
current-gauge field interaction term. While preserving  gauge invariance, this modified interaction accounts for
the vacuum dielectric properties at different length scales.
This is a re-interpretation of the original Lee-Wick model,  leading to a linear behavior at short distances. \\
On the other hand, to reproduce confinement, one needs a linear potential at \emph{large} distances.
 At the phenomenological level, the Cornell potential, reproducing a linear behavior, is  currently adopted 
to fit the quark/anti-quark bound  state spectrum \cite{Eichten:1978tg}.\\

\begin{equation}
 g_sV_C\left(\, r \,\right) = -\frac{4}{3}\frac{\alpha_s}{r} + \sigma r \ ,
\label{c1}
\end{equation}

The first term, which dominates the short distances behavior, is a Coulomb potential with the fine structure constant 
being  replaced by  the \emph{strong coupling}  constant $\alpha_s= g^2_s/4\pi$.
On the other hand, the second term provides a constant confining force. The constant $\sigma$ 
represents the \emph{tension} of the color flux tube connecting the  quark/anti-quark pair.\\
 In $QCD$, it is not at all evident that confinement is actually an Abelian ``~\emph{phenomenon}~'' as shown in
\cite{Luscher:1978rn,Kondo:1997pc,Kondo:1997kn,Kondo:2014sta}. 
Nevertheless, isolating the commuting
degrees of freedom in a Yang-Mills gauge theory is a  complicated task and  recovering a linear
potential is technically even more difficult. This is a good motivation to look for an equivalent model 
in a simpler framework. \\

Following the results of the previous Section, we introduce a possible redefinition of the Lee-Wick 
action which exchanges the short and long distance behavior. The new Lagrangian is

\begin{equation}
 L = -\frac{1}{4}F_{\mu\nu} \frac{-\partial^2}{-\partial^2 + m^2} F^{\mu\nu} - e J^\mu A_\mu
\label{ab}
\end{equation}

It is known that  confinement kicks in at an energy  determined by the dynamically generated $QCD$ 
scale $\Lambda_{QCD}\approx 200 MeV$. To simulate confinement in the Abelian model (\ref{ab})
we take $m\sim \Lambda_{QCD}$.\\
In order to calculate the new potential we start as in (\ref{fourier})

\begin{eqnarray}
 \phi\left(\,r\,\right) &&= -\frac{e}{\varepsilon_0}\int \frac{d^3k}{\left(\,2\pi\,\right)^3}\frac{1}{\vec{k}^{\,2}}
\frac{\vec{k}^{\,2}+m^2}{\vec{k}^{\,2}} e^{i\vec{k}\cdot \vec{r}}\ ,\nonumber\\
&&=-\frac{e}{\varepsilon_0}\int \frac{d^3k}{\left(\,2\pi\,\right)^3}\frac{1}{\vec{k}^{\,2}}
\left(\, 1 +\frac{m^2}{\vec{k}^{\,2}}\,\right)e^{i\vec{k}\cdot \vec{r}}
\label{noso}
\end{eqnarray}

Again, the first term in (\ref{noso}) gives the Coulomb part of the potential 

\begin{equation}
 \int \frac{d^3k}{\left(\,2\pi\,\right)^3}\frac{1}{\vec{k}^{\,2}}e^{i\vec{k}\cdot \vec{r}}= \frac{1}{4\pi r}
\end{equation}

This part dominates the short-distances behavior.\\
 The second term leads to a linearly rising potential

\begin{eqnarray}
 m^2\int \frac{d^3k}{\left(\,2\pi\,\right)^3}\frac{1}{\left(\,\vec{k}^{\,2}\,\right)^2}e^{i\vec{k}\cdot \vec{r}} &&=
\frac{m^2}{\left(\,2\pi\,\right)^3}\int_0^\infty ds s  \int d^3k e^{-s\vec{k}^{\,2} }e^{i\vec{k}\cdot \vec{r}}\ ,
\nonumber\\
&&=\frac{m^2}{8\pi^{3/2}} \int_0^\infty ds s^{-1/2} e^{-r^2/4s}\ ,\nonumber\\
&&=\frac{m^2}{8\pi^{3/2}}\, \frac{r}{2}\, \Gamma\left(\, -1/2\,\right)= - \frac{m^2}{8\pi}r
\end{eqnarray}

The complete result for $\phi$ is encompassed in

\begin{equation}
 e\phi\left(\, r\,\right)= - \frac{e^2}{4\pi\varepsilon_0 }\frac{1}{r}\left(\, 1 - \frac{m^2}{2} r^2\,\right)
\end{equation}

The correspondence with the Cornell potential (\ref{c1}) is established through the following identifications

\begin{eqnarray}
 && e^2 \Longleftrightarrow \frac{4}{3} \alpha_s\ ,\\
 && m^2 e^2 \Longleftrightarrow \frac{4}{3} \alpha_s \Lambda^2_{QCD}\ ,\\
 && \sigma = \frac{1}{4\pi\varepsilon_0}\frac{4}{3} g^2_s\Lambda^2_{QCD}
\end{eqnarray}
 
It is interesting to notice the singular behavior of the ``~color~'' dielectric constant $\varepsilon\left(\, r\,\right)$ 
as $r$ approaches the "~\emph{critical}~" distance $r^\ast = \sqrt{2}/ m $

\begin{equation}
 \varepsilon\left(\, r\,\right)=  \frac{  \varepsilon_0 }{1 - m^2 r^2/2  }
\end{equation}

The Coulomb region corresponds to $0\le r < r^\ast$, where the running dielectric constant 
$\varepsilon\left(\, r\,\right)$ is positive and diverges at the critical distance $r^\ast$.
Beyond this point begins the "~large distances~" region $r>r^\ast $, where $\varepsilon\left(\, r\,\right)$ is negative 
 and vanishes as $r\to \infty$. This behavior can be surprising at first, but its possible  physical
explanation could be  that it is 
impossible to separate two charges beyond a distances at which the creation of a new pair is energetically favorable.
Thus, the infinite discontinuity in $r=r^\ast$ marks a "~phase transition~" between the Coulomb
and the confining phase.

\subsection{Gaussian source}

As shown in the previous Section, a suitable modification of the Maxwell action leads to the Cornell confining
potential at large distances. However, the Coulomb potential still emerges at short distances  with the well-known 
singular behavior at the origin. A similar pattern is encountered both in the Newtonian gravitational potential  
and in its relativistic extension. In General Relativity this problem acquires a particular relevance since it leads 
to the, so-called, \emph{curvature singularity} issue. This invalidates  basic assumptions of the space-time 
being a smooth, differentiable,  manifold.  On a physical ground, even in the case when the singularity is 
``~\emph{hidden}~'' by an event horizon,
the very existence of the singularity prevents a consistent description of the final stage of black-hole evaporation.\\
Recently, this problem has been successfully solved by the natural assumption that the matter source of the gravitational
field cannot be point-like. It has been replaced by modeling a ``~particle~'' as a Gaussian matter/energy distribution.
This choice was motivated by the known property of Gaussian states in Quantum Mechanics, as being the best approximation
of a classical, point-like object  
\cite{Nicolini:2005vd,Ansoldi:2006vg,Ansoldi:2008jw,Spallucci:2009zz,Nicolini:2008aj,Nicolini:2009gw,Spallucci:2017aod}.\\
Based on this experience, we would like to apply the same idea to obtain a \emph{regular} Cornell potential. The physical
motivation is that a confining potential is instrumental to describe the properties of hadrons, which are, surely, not
point-like objects.\\ 
Thus, we start  by replacing the Dirac delta-function by a Gaussian of a finite width $l_0$ 

\begin{equation}
 \delta\left(\, \vec{r}\,\right)\longrightarrow \frac{1}{\left(\, 2\pi l_0^2\,\right)^{3/2} } e^{-r^2/4l_0^2}
\end{equation}

The form of the charge distribution turns out to have a particularly simple form  in the momentum space

\begin{equation}
 \rho\left(\, \vec{k}\,\right) =  e^{-\vec{k}^2 l^2_0}\ , \label{rog}
\end{equation}

 At this point,  it is important to
mention that we have two length scales in the theory:  $l_0$ and  $1/m$. The former is needed to
regularize the theory at short-distances and is assumed to be $l_0 << 1/m$, while the latter gives a dominant 
contribution at large distances.  \\
Now, we shall repeat the steps from the previous Sections using (\ref{rog}) in (\ref{noso}) to obtain

\begin{eqnarray}
 \phi\left(\,r\ ,l_0\,\right) &&= -\frac{e}{\varepsilon_0}\int \frac{d^3k}{\left(\,2\pi\,\right)^3}\frac{1}{\vec{k}^{\,2}}
\frac{\vec{k}^{\,2}+m^2}{\vec{k}^{\,2}} e^{i\vec{k}\cdot \vec{r}} e^{-l_0^2\,\vec{k}^{\,2} }\ ,\nonumber\\
&&=-\frac{e}{\varepsilon_0}\int \frac{d^3k}{\left(\,2\pi\,\right)^3}\frac{1}{\vec{k}^{\,2}}
\left(\, 1 +\frac{m^2}{\vec{k}^{\,2}}\,\right)e^{i\vec{k}\cdot \vec{r}} e^{-l_0^2\,\vec{k}^{\,2} }\ ,\nonumber\\
&&\equiv \phi\left(\,r\ ,l_0\ ; m=0\,\right)+\phi\left(\,r\ ,l_0\ ; m\,\right)
\label{fiint}
\end{eqnarray}

Let us begin by  evaluating the $m$-independent part of the integral  (\ref{fiint}):

\begin{eqnarray}
\phi\left(\,r\ ,l_0\ ; m=0\,\right)&&= -\frac{e}{\varepsilon_0}\int \frac{d^3k}{\left(\,2\pi\,\right)^3}\frac{1}{\vec{k}^{\,2}}e^{i\vec{k}\cdot \vec{r}} e^{-l_0^2\,\vec{k}^{\,2} }
\ ,\nonumber\\ 
&&=-\frac{e}{\varepsilon_0}\int_0^\infty ds \int \frac{d^3k}{\left(\,2\pi\,\right)^3}
e^{i\vec{k}\cdot \vec{r}} e^{-\left(\, l_0^2+s\,\right)\,\vec{k}^{\,2} }\ ,\nonumber\\
&&=-\frac{e}{\varepsilon_0}\frac{1}{8\pi^{3/2}}\int_{l_0^2}^\infty d\tau \tau^{-3/2}\, e^{-r^2/4\tau}\ , \tau=s + l_0^2\ ,\\
&&=-\frac{e}{\varepsilon_0}\frac{1}{4\pi\sqrt{\pi}}\frac{1}{r}\gamma\left(\, \frac{1}{2}\ ; \frac{r^2}{4l_0^2}\,\right)\ ,
\end{eqnarray}

The asymptotic behavior of $\phi\left(\,r\ ; m=0\,\right)$ is ( see Appendix ):

\begin{eqnarray}
\phi\left(\,r\ ,l_0\ ; m=0\,\right) &&\sim -\frac{e}{4\pi\varepsilon_0}\frac{1}{r}\ ,\qquad r\longrightarrow \infty\ ,\\
&&\sim -\frac{e}{\varepsilon_0}\frac{1}{4\pi^{3/2}\, l_0} \,\left(\, 1-\frac{r^2}{4l^2_0}+\dots \,\right) 
\ ,\quad r<< l_0 \label{shortd}
\end{eqnarray}

\begin{equation}
 \phi\left(\,r=0\ ,l_0\ ; m=0\,\right)=-\frac{e}{4\pi^{3/2}\varepsilon_0\, l_0}\label{potzero}
\end{equation}

Equations (\ref{shortd}) and (\ref{potzero}) show that the $m$-independent part of the total potential corresponds
to the regular Coulomb part.  The width $l_0$ of the Gaussian source turns out to be a natural cut-off eliminating the
singularity in $r=0$.\\
On the other hand, the $m$-dependent part of the potential is crucial to obtain a linear behavior at large distances.

\begin{eqnarray}
\phi\left(\,r\ ,l_0\ ; m\,\right)&=&-\frac{e}{\varepsilon_0}\,
 \int \frac{d^3k}{\left(\,2\pi\,\right)^3}\frac{m^2}{\left(\,\vec{k}^{\,2}\,\right)^2}\,
e^{i\vec{k}\cdot \vec{r}} e^{-l_0^2\,\vec{k}^{\,2} }\ ,\nonumber\\
=&-&\frac{e}{\varepsilon_0}\,\frac{m^2}{8\pi^{3/2}}\int_0^\infty ds \frac{s}{\left(\, s +l_0^2\,\right)^{3/2}}e^{-r^2/4\left(\, s + l_0^2\,\right)}
\ ,\nonumber\\
=&-&\frac{e}{\varepsilon_0}\,\frac{m^2}{8\pi^{3/2}}\int_{l_0^2}^\infty \frac{d\tau}{\tau^{1/2}}\left(\, 1 - \frac{l_0^2}{\tau}\,\right)\,
e^{-r^2/4\tau} ,\tau= s +l_0^2 \ \\
&& \equiv \phi\left(\,r\ ,l_0=0\ ; m\,\right)+\widetilde{\phi}\left(\,r\ ,l_0\ ; m\,\right)\label{contribution}
\end{eqnarray}

As before, we compute the two contributions in (\ref{contribution}) separately starting with 
$\widetilde{\phi}\left(\,r\ ,l_0\ ; m\,\right)$:

\begin{eqnarray}
 \widetilde{\phi}\left(\,r\ ,l_0\ ; m\,\right)&&= -\frac{e}{\varepsilon_0}\,\frac{m^2l^2_0}{8\pi^{3/2}}\int_{l_0^2}^\infty \frac{d\tau}{\tau^{3/2}}e^{-r^2/4\tau} =\frac{m^2l^2_0}{4\pi^{3/2}}
\frac{1}{r}\gamma\left(\, \frac{1}{2}\ ; \frac{r^2}{4l_0^2}\,\right)\, \, \, \ \ \\
&& \sim -\frac{e}{\varepsilon_0}\, \frac{m^2l_0}{4\pi^{3/2}} 
\left(\, 1 -\frac{r^2}{6l_0^2}+\dots\,\right)\ ,\quad r<< l_0
\end{eqnarray}

Then, we calculate the remaining $l_0$-independent part

\begin{eqnarray}
  \phi\left(\,r\ ,l_0=0\ ; m\,\right)
&&=-\frac{e}{\varepsilon_0}\,\frac{m^2}{8\pi^{3/2}}\frac{r}{2}\int_0^{r^2/4l_0^2}\frac{du}{u^{3/2}} e^{-u} \ ,\qquad u=r^2/4\tau\ ,\nonumber\\
&&=-\frac{e}{\varepsilon_0}\,\frac{m^2}{8\pi^{3/2}}\frac{r}{2}\, \gamma\left(\, -1/2\ ; r^2/4l_0^2\,\right)\ ,\\
&&\sim \frac{e m^2}{8\pi\,\varepsilon_0}\,r\ ,\qquad r>> l_0 \label{linear}
\end{eqnarray}

Equation (\ref{linear}) exhibits the expected linear behavior at large distances.\\
Therefore, the complete potential is given by
{\small
\begin{eqnarray}
&& \phi\left(\,r\ ,l_0\ ; m\,\right)= \nonumber\\
&-& \frac{e}{4\pi^{3/2}\,\varepsilon_0 }\frac{1}{r}\left[\,
\left(\, 1 -m^2 l_0^2\,\right)\, \gamma\left(\, 1/2\ ; r^2/4l_0^2\,\right) 
+\frac{m^2}{4} r^2 \gamma\left(\, -1/2\ ; r^2/4l_0^2\,\right) \,\right]\, \, \, \, \, \, \, \, 
\end{eqnarray}
}

As it was anticipated, the spread of the source modifies only the short-distances behavior of the potential rendering
 the potential finite at the origin:

\begin{equation}
 \phi\left(\,r=0\,\right)=- \frac{e}{4\pi^{3/2}\varepsilon_0\, l_0 }\left(\, 1 -m^2 l_0^2\,\right)
\approx - \frac{e}{4\pi^{3/2}\varepsilon_0\, l_0 } \label{last}
\end{equation}

Our previous assumption $m\, l_0 << 1$ assures the 
attractive character of the interaction near the origin.

\begin{figure}[h!]
\begin{center}
\includegraphics[height=8cm]{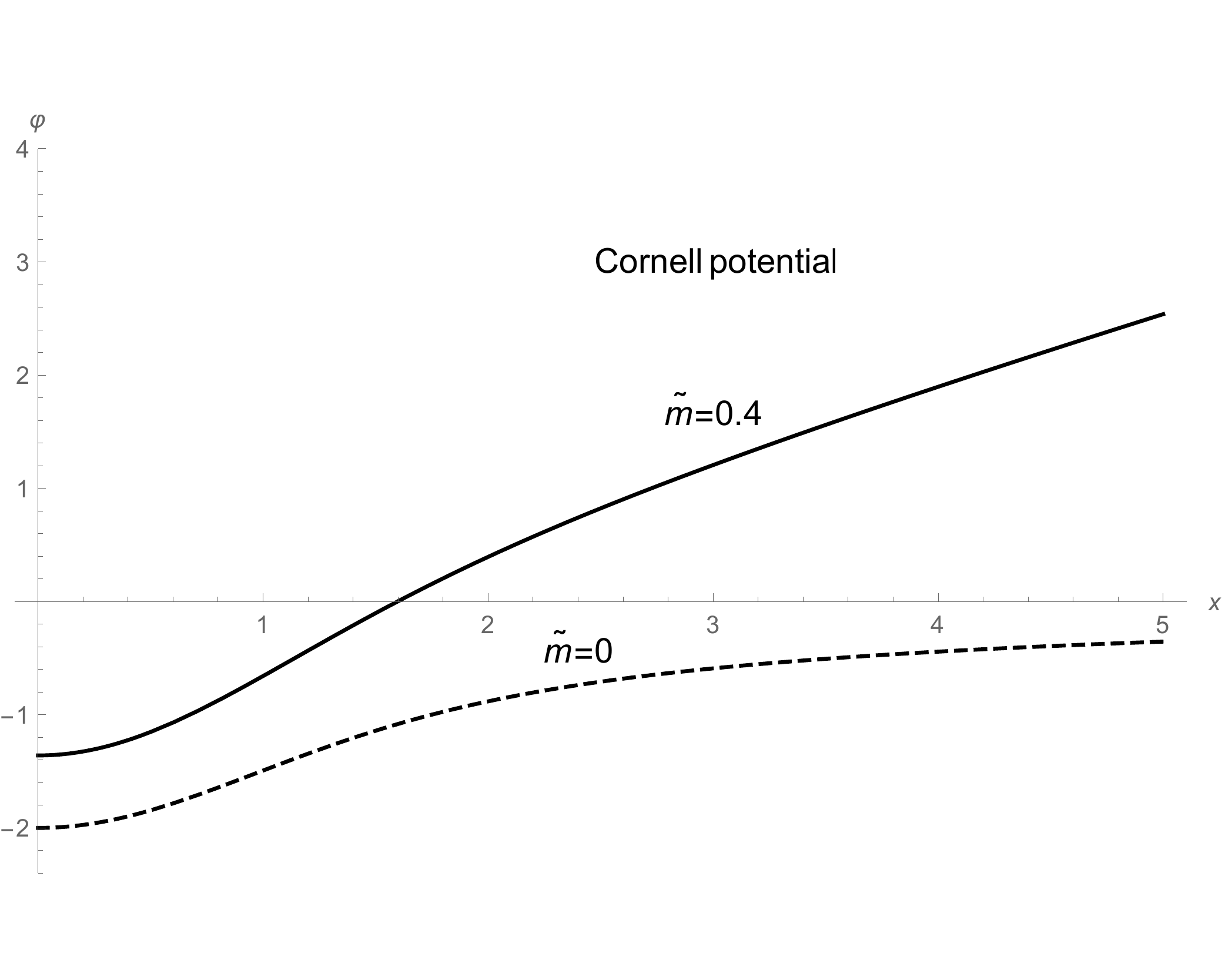}
\caption{Plot of the potential in eq.(\ref{last}) for $\widetilde{m}\equiv  ml_0=0$ ( regular Coulomb potential at short distances ), 
and $\widetilde{m}\equiv ml_0=0.4$, i.e. regular Cornell potential.}
\label{due}
\end{center}
\end{figure}

\section{Conclusions}

At the beginning of this work we have briefly reviewed the  Lee-Wick modification of ordinary electrodynamics 
leading to a linear electrostatic potential at short distances, while preserving the Coulomb form at large distances. 
We have also shown that the higher derivative  correction to the kinetic term can be alternatively shifted 
into the interaction
term. In this way, the kinetic term retains its canonical form and all the problems, at the quantum level, related to
the presence of higher order derivatives are avoided. The new interaction term can be seen either as a non-point-like
source of the field or as an effective dielectric constant depending on the distances from the source charge. 
In the latter interpretation the source remains point-like while the electric properties of the surrounding vacuum 
change with distances.\\
From this vantage point, one can further proceed to define a suitably modified interaction term. This leads
to the exchange between the short and long
distances behavior of the original Lee-Wick model. In this way, one obtains the Cornell potential
between electric charges. Within this framework, one simulates the behavior  expected to take place for color
charges in Quantum Chromodynamics. Without pretending to construct a phenomenologically 
relevant model we, nevertheless, hope this work gives useful insights into the more realistic, but so far unsolved, 
case of Yang-Mills gauge theories.\\
In the last part of the work, we  transferred the results, obtained in General Relativity, in order to remove
the singular behavior of the Cornell potential in $r=0$. For this purpose, we have replaced point-like charges
by distributed ones. The final result is a modified Cornell potential which remains linear at large
distances but turns out to be finite in $r=0$.

\section{Useful formulas}

The Fourier integrals are computed using the Schwinger formula:
\begin{equation}
 \frac{1}{\left(\,\vec{k}^{\,2}\,\right)^\alpha }=\frac{1}{\Gamma\left(\,\alpha\,\right)}
\int_0^\infty ds\,s^{\alpha-1} e^{-s\,\vec{k}^{\,2} }
\end{equation}

Gauss Integral:

\begin{equation}
 \mathbf{\int d^n x \, e^{-\frac{1}{2} x\, A\, x} e^{B\cdot x}=e^{\frac{1}{4} B A^{-1}B}
\left(\,\frac{\pi}{det\left(\, A \,\right)}\,\right)^{n/2}}
\end{equation}
 
Lower incomplete Gamma Function definition

\begin{equation}
 \gamma\left(\, \alpha\ ; z\,\right)\equiv \int_0^z dt\, t^{\alpha-1} e^{-t}\ ,\qquad Re\,\alpha >0
\end{equation}

The asymptotic behavior of $\gamma\left(\, \alpha\ ; z\,\right)$ at small $z$ is 

\begin{equation}
  \gamma\left(\, \alpha\ ; z\,\right)\sim \frac{1}{\alpha} z^\alpha
\end{equation}

Upper incomplete Gamma Function definition

\begin{equation}
 \Gamma\left(\, \alpha\ ; z\,\right)\equiv \int_z^\infty dt\, t^{\alpha-1} e^{-t}
\end{equation}

\begin{equation}
 \gamma\left(\, \alpha\ ; z\,\right)+\Gamma\left(\, \alpha\ ; z\,\right)=\Gamma\left(\, \alpha\,\right)\ ,\qquad
\alpha \Gamma\left(\, \alpha\,\right)=\Gamma\left(\, \alpha +1\,\right)
\end{equation}

where $\Gamma\left(\, \alpha\,\right)$ is the Euler Gamma Function.

\begin{equation}
 \Gamma\left(\, \frac{1}{2}\,\right)=\sqrt\pi
\end{equation}

\end{document}